\documentclass{aa}
\usepackage[utf8]{inputenc}
\usepackage{graphicx}
\usepackage{txfonts}
\usepackage{verbatim}
\usepackage{siunitx}
\usepackage{hyperref}
\hypersetup{colorlinks, allcolors=blue}
\usepackage{subcaption}
\usepackage{comment}
\usepackage[labelfont=bf]{caption}
\usepackage{tablefootnote}
\usepackage[flushleft]{threeparttable}
\usepackage{caption}
\usepackage{amsmath}
\usepackage{relsize}
\usepackage{multirow}
\usepackage{mathtools}

\usepackage[switch]{lineno}
\begin{document}

\title{Peering above the clouds of the warm Neptune GJ~436\,b with CRIRES+}
\subtitle{} 
\titlerunning{Atmospheric characterization of GJ~436b with CRIRES+}

\author{Natalie Grasser\inst{1} \and Ignas A. G. Snellen\inst{1} \and Rico Landman\inst{1} \and Darío González Picos\inst{1} \and Siddharth Gandhi\inst{2,3,1}}
\institute{Leiden Observatory, Leiden University, Postbus 9513, 2300 RA, Leiden, The Netherlands \\
\email{grasser@strw.leidenuniv.nl} \and Department of Physics, University of Warwick, Coventry CV4 7AL, United Kingdom \and Centre for Exoplanets and Habitability, University of Warwick, Gibbet Hill Road, Coventry CV4 7AL, United Kingdom}
\date{}

\abstract{Exoplanets with masses between Earth and Neptune are amongst the most commonly observed, yet their properties are poorly constrained. Their transmission spectra are often featureless, which indicate either high-altitude clouds or a high atmospheric metallicity. The archetypical warm Neptune GJ~436\,b is such a planet showing a flat transmission spectrum in observations with the \textit{Hubble} Space Telescope (HST).} 
{Ground-based high-resolution spectroscopy (HRS) effectively probes exoplanet atmospheres at higher altitudes and can therefore be more sensitive to absorption coming from above potential cloud decks. In this paper we aim to investigate this for the exoplanet GJ~436\,b.} 
{We present new CRIRES+ HRS transit data of GJ~436\,b. Three transits were observed, but since two were during bad weather conditions, only one transit was analyzed. The radiative transfer code petitRADTRANS was used to create atmospheric models for cross-correlation and signal-injection purposes, including absorption from H$_2$O, CH$_4$, and CO.}
{No transmission signals were detected, but atmospheric constraints can be derived. Injection of artificial transmission signals indicate that if GJ~436\,b would have a cloud deck at pressures $P$>10~mbar and a <300$\times$ solar metallicity, these CRIRES+ observations should have resulted in a detection.} 
{We estimate that the constraints presented here from one ground-based HRS transit are slightly better than those obtained with four HST transits. Combining HRS data from multiple transits is an interesting avenue for future studies of exoplanets with high-altitude clouds.}


\maketitle

\section{Introduction}

The warm Neptune-sized exoplanet GJ~436\,b is a prime example of an intermediate mass planet with sizes between those of Earth and Neptune. Observational surveys with the Kepler Space Telescope and the HARPS spectrograph have revealed that such exoplanets are amongst the most common (e.g., \citealt{Borucki2011,Mayor2011,Howard2012,Batalha2013,Fressin2013}), yet their characteristics are poorly understood. The mean densities of such planets are compatible with a large variety of compositions, ranging from rocky cores with thick hydrogen envelopes to water-rich planets with steamy atmospheres (e.g., \citealt{ElkinsTanton2008,Figueira2009,MillerRicci2010}). Close-in sub-Neptune planets are thought to either form beyond the ice line and migrate inwards, or form close-in via coagulation of inward-drifting pebbles (e.g., \citealt{Bean2021}). Determining the atmospheric composition of these planets could resolve degeneracies of interior structure models and shed light on their formation (e.g., \citealt{Madhusudhan2023, Benneke2024}). Considering their frequency, it would significantly contribute to our understanding of how planets form and evolve.

GJ~436\,b was the first Neptune-sized exoplanet ($M_\text{p}=1.3~M_{\text{Nep}}$, $R_\text{p}=1.1~R_{\text{Nep}}$) discovered, making it the lowest mass exoplanet at the time of its discovery, and therefore it has been extensively studied. Orbiting the M2.5V-type star GJ~436 ($M_{*}=0.4~M_{\odot}$, $\text{age}>3$~Gyr), with a distance of $d=10.23$~pc from Earth, the planet was first discovered by \cite{Butler2004} through radial velocity variations of its host star, and was eventually also revealed to be a transiting planet by \cite{Gillon2007}. It has an orbital period of $P=2.644$~days, a semi-major axis of $a=0.028$~AU (at 14 stellar radii) and an eccentricity of $e=0.145$. Due to its close-in orbit, its equilibrium temperature is estimated to be around $650-800$~K, making it a ``warm Neptune'' (e.g., \citealt{Torres2008,Lanotte2014,Knutson2014,Turner2016,Lothringer2018}).

Before atmospheric measurements were available, estimates were made on the possible composition of GJ~436\,b based on its radius and mass. Using planetary structure and formation models, \cite{Figueira2009} derived that the planet is unlikely to consist of more than 30\% hydrogen-helium gases (akin to a small Jovian planet) or 90\% rock and iron (akin to a super Earth). Instead, they suggest more intermediate properties, with some degree of heavy-element enrichment, as well as significant amounts of water in each formation scenario, making its composition differ significantly to that of Neptune. Both \cite{Figueira2009} and modeling by \cite{Nettelmann2010} imply that some form of hydrogen-helium envelope is required to be compatible with the observed radius.

First atmospheric data were obtained from the planet's thermal emission at secondary eclipse by the \textit{Spitzer} Space Telescope. Among traces of H$_2$O and CO$_2$, \cite{Stevenson2010} find that the planet has a high CO abundance and a significant CH$_4$ deficiency relative to equilibrium models for a hydrogen-dominated atmosphere, which predict that CH$_4$ should be the dominant carbon-bearing species, while a reanalysis of this data by \cite{Lanotte2014} finds consistent but lower significance of these results. These observations could be explained by a combination of high metallicity and disequilibrium chemistry such as vertical mixing, though likely not to a sufficient extent \citep{Madhusudhan2011,Line2011}. A modeling approach by \cite{Moses2013} suggests that moderate-to-high atmospheric metallicities ($230-2000\times$ solar) could create a CO-rich, CH$_4$-depleted atmosphere and would also be consistent with the heavy-element enrichment required to explain the planet's mass and radius. 

Transit observations of GJ~436\,b by the \textit{Hubble} Space Telescope (HST) yield a flat transmission spectrum without any molecular absorption features \citep{Knutson2014}. Their best-fitting atmospheric models have either high-altitude clouds or a high metallicity. At high metallicities, the scale height of the atmosphere is reduced, producing smaller absorption features in transmission spectra, whereas at low metallicities, a cloud deck at pressures below 10~mbar would produce a similarly flat spectrum. Further studies based on modeling and observations are consistent with a metal-rich ($100-1000\times$ solar), though the source of a sufficiently large CH$_4$ depletion is still unclear \citep{Lanotte2014,Agundez2014,Morley2017,Lothringer2018}. 

In addition, \cite{Hu2015} propose that a helium-dominated atmosphere is compatible with the observed emission and transmission spectra, which could be created through hydrodynamic escape of hydrogen due to the high stellar irradiation caused by the planet's proximity to its star. With the depletion of hydrogen, the dominant carbon-bearing molecule can no longer be CH$_4$, but would instead be CO. A metal-rich atmosphere moderately high in hydrogen could also cause CO to dominate over CH$_4$, since CH$_4$ and H$_2$O would be competing for hydrogen. To create the observed featureless transmission spectrum, an aerosol layer at pressures of approximately 1~mbar would be necessary in both of these scenarios. Significant atmospheric escape has been found using Ly-$\alpha$ observations, which show that GJ~436\,b is indeed enshrouded and trailed by a large cloud of hydrogen \citep{Ehrenreich2015}. Despite the expected helium triplet absorption as a further tracer of atmospheric evaporation, observations resulted in a non-detection \citep{Nortmann2018}, which could be explained by a He/H ratio three times smaller than solar \citep{Rumenskikh2024}. Despite extensive efforts to characterize GJ~436\,b's atmosphere, many open questions remain due to ambiguous observational data. In fact, featureless transmission spectra for this population of planets is a very common phenomenon (e.g., \citealt{Bean2010, Kreidberg2014,Ehrenreich2014,Knutson2014}).

These long-standing challenges call for high-resolution observations to characterize exoplanetary atmospheres. Since the signal of the planet is orders of magnitude smaller than that of the star and/or tellurics, its molecular spectral features are deeply buried in the noise. In order to retrieve a planetary signal, the signal from all of the individual spectral lines are combined by cross-correlating with a template of the expected absorption lines. Since the first robust detection of a molecule in an exoplanet atmosphere \citep{Snellen2010} using this method, high-resolution cross-correlation spectroscopy (HRCCS) has proven to be a powerful technique for characterizing exoplanet atmospheres using ground-based telescopes. So far, it has been successful in identifying molecular species, atmospheric structures, spin rotation, and day-to-night winds through transmission and thermal emission spectroscopy of transiting and non-transiting exoplanets (e.g., \citealt{Brogi2012,Brogi2016,Birkby2017,Kesseli2020,Hoeijmakers2020,Ehrenreich2020})

\cite{Gandhi2020} predict that high-resolution spectroscopy (HRS) using cross-correlation techniques could partially break this degeneracy by distinguishing between the transmission spectrum of a hydrogen-dominated, cloudy atmosphere, and that of a clear atmosphere of high metallicity. A similar conclusion on the potential of HRS is independently obtained by \cite{Hood2020}. In contrast to low-resolution spectroscopy, which constrains species through their broad-band absorption features, HRS resolves individual lines, through which it can distinguish between different species and effectively probe higher in the atmosphere.

An ideal instrument for this task is CRIRES+ (upgraded CRyogenic high-resolution InfraRed Echelle Spectrograph) on the Very Large Telescope (VLT) at the Paranal Observatory in Chile (e.g., \citealt{Kaeufl2004,Follert2014, Dorn2014M}). It has recently been upgraded to a cross-dispersed spectrograph, increasing the instantaneous spectral coverage by up to a factor of ten. Thanks to its  high spectral resolution of $\mathcal{R}=85,000-100,000$, CRIRES+ is sensitive to the strong spectral features that form at low pressures above potential clouds, while lower-resolution spectroscopy is more sensitive to the combined absorption from the numerous weaker lines originating deeper in the atmosphere. 

In this work, we present our analysis of new CRIRES+ transit spectroscopy of the GJ~436 system. We focus on the detectability of planetary H$_2$O, as it is among the most well observed spectroscopically active molecules and can exist in a large range of atmospheric compositions and temperatures (e.g., \citealt{Madhusudhan2012}). Assuming chemical equilibrium, our simulations indicate that H$_2$O would be the easiest molecule to detect in all available wavelength settings. In Section~\ref{sec:obs}, we describe our observations and their reduction. We present our data analysis in Section~\ref{sec:methods}, which includes the generation of model atmospheres and an outline of our cross-correlation detection pipeline. In Section~\ref{sec:results}, we present our main results and discuss them in Section~\ref{sec:discuss}, concluding our work in Section~\ref{sec:concl}.

\section{Observations} \label{sec:obs}

\begin{figure*}
    \centering
    \includegraphics[width=\textwidth]{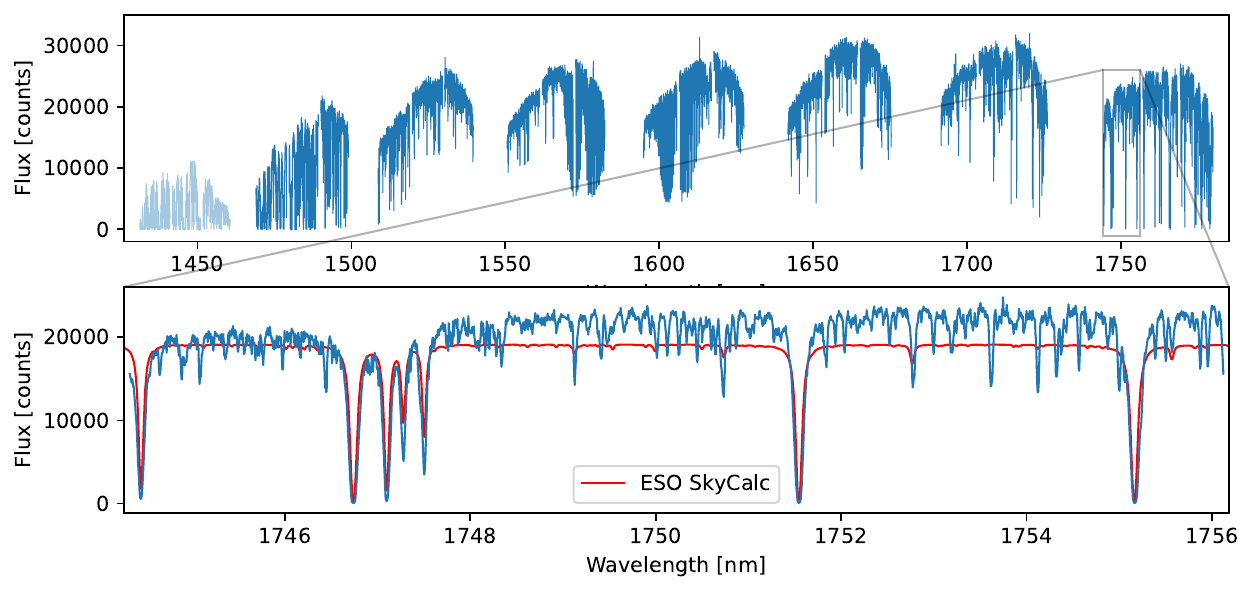}
    \caption{Observed spectrum of the GJ~436\,b system during mid-transit. Top panel: All 8 spectral orders. Due to heavy contamination by tellurics, we discard the bluest order shown as semi-transparent. Bottom panel: Zoom-in on the reddest spectral order, compared with the telluric absorption model from the ESO SkyCalc tool (red).}
    \label{fig:original_spec}
\end{figure*}

We acquired new transit spectroscopy data of GJ~436 using the upgraded CRIRES+ instrument on the VLT at Cerro Paranal. Since the change in planet radial velocity during the transit is relatively small ($\sim$10 km/s), it is important to choose the date of observation in such way that the barycentric velocity of the Earth results in a significant offset between the anticipated planet water signal and that from the Earth atmosphere. 
The observations were conducted on the nights starting on January 22, January 30, and March 24, 2023 (PI: R. Landman), using individual exposure times of 60 seconds. Unfortunately, for the second and third nights, the adaptive optics system could not be used due to high humidity, resulting in up to eight times less flux. Therefore, we only used the data from January 22 for our analysis. A fourth night was planned, but problematic weather conditions prevented this. The observations of the January 22 transit covered orbital phases between -0.01081 and 0.01421, with an airmass of 1.8 at the start of the transit to 1.6 at the end. The conditions during the observations were good, with an average seeing of $\sim$0.65". 

Our simulations in the J-, H-, and K-band resulted in the highest S/N when using the H1567 wavelength setting. Therefore, the H1567 setting was used to maximize the potential signals from water vapor and methane, covering a wavelength range of $1490.01-1779.97$ nm over 8 spectral orders that are each sampled by 3 detector arrays. We used the 0.2" slit, resulting in a spectral resolution of approximately $\mathcal{R}=100,000$. To enable a sky-background subtraction, the observations were conducted in two nodding positions separated by 5", with 45 exposures each. No jittering was employed to keep the systematics constant over the course of the transit. We reduced the data using the official ESO CR2RES pipeline (Version 1.4.0) \footnote{\url{https://ftp.eso.org/pub/dfs/pipelines/instruments/cr2res/cr2re-pipeline-manual-1.4.0.pdf}} together with the Python wrapper pycrires \citep{Stolker2023}. To determine suitable parameters for the reduction, several different variations were tested for the spectral extraction, using the retrieved signal-to-noise ratio (S/N) of injected model atmospheres as a measure of quality (determined by taking the difference of the CCFs with and without the injection and dividing by the standard deviation of the non-injected run, see Section~\ref{sec:methods}). Doing so, we found that using the automatically calculated extraction height, a swath width of 400 pixels, not subtracting the nolight rows nor the interorder column worked best. The wavelength solution was initially obtained using the calibration files from the ESO pipeline. We subsequently fine-tuned this wavelength solution by maximizing the cross-correlation signal between the observed spectrum and a telluric template, as described in detail in \citet{Landman2023}. We found that only small corrections to the initial wavelength solution were required, with offsets of maximally 0.01 nm. We estimate from the CCFs with the telluric templates that we reach a precision of about 0.006~nm. The wavelength solution was checked by visually comparing the obtained spectrum with a telluric model.

Figure~\ref{fig:original_spec} shows the observed spectrum from exposure 20 in nodding position B, with the top panel containing all 8 spectral orders. It was taken approximately mid-transit. The bluest telluric-dominated order, shown semi-transparent, is removed from our analysis. With this, we achieve a higher S/N of injected model atmospheres. The bottom panel shows a zoom-in on the reddest order. Using the ESO SkyCalc tool (Version 2.0.9, \citealt{Noll2012,Jones2013}), a model of the telluric absorption at the time of the transit was created, which we show in red. The precise overlap of the telluric lines in the model and our data visually confirms the accuracy of our wavelength solution. Aside from the tellurics, the stellar absorption lines are also clearly visible. The top panel in Figure~\ref{fig:process} shows the observed spectra of all exposures in nodding position B of a part of the reddest order.

\section{Data analysis} \label{sec:methods}

\subsection{Model atmospheres} \label{subsec:temp}

Models of the planet transmission spectra were created using petitRADTRANS (pRT, \citealt{Molliere2019}), which is a Python package for calculating transmission and emission spectra of exoplanets through radiative transfer. These model atmospheres are used as templates in our cross-correlation algorithm (see Section~\ref{subsec:pipeline}). We assumed a H$_2$-He dominated atmosphere, which is required to explain the observed radius of the planet \citep{Figueira2009,Nettelmann2010}. For these species, Rayleigh scattering and collision induced absorption are included. Furthermore, we assume a temperature of 700~K (e.g., \citealt{Turner2016}). Since high-resolution transit spectroscopy mainly probes low pressures and is less sensitive to absolute abundances or changes in the pressure-temperature profile, using isothermal profiles is a sufficient approximation. Clouds are modeled using the gray cloud deck implementation from pRT. %Further continuum opacity sources include collision induced absorption from H$_2$-H$_2$ and H$_2$-He, as well as Rayleigh scattering from H$_2$ and He.

In order to investigate the presence of different molecules, we created separate templates for H$_2$O, CH$_4$ and CO using the main isotopolog default line lists from the HITEMP database \citep{HITEMP2010,HITEMP2020_CH4}. Assuming a solar C/O ratio of 0.55 (e.g., \citealt{Asplund2009}), the abundances of these molecules and the mean molecular weight (MMW) were calculated using pRT's poor\_mans\_nonequ\_chem subpackage, which interpolates the abundances of prominent species from a chemical equilibrium table. For each of these molecular species, we generated a grid of templates with varying metallicities, ranging from solar to 1000$\times$ solar, and cloud deck altitudes, ranging from 0.1~mbar to 1 bar, spaced evenly on a log-scale, resulting in a total of 399 templates for each species. With this, we aim to place constraints on the metallicity of the atmosphere and examine the presence of a potential high altitude cloud deck, as suggested by \cite{Knutson2014}.

\begin{figure}
    \centering
    \includegraphics[width=\columnwidth]{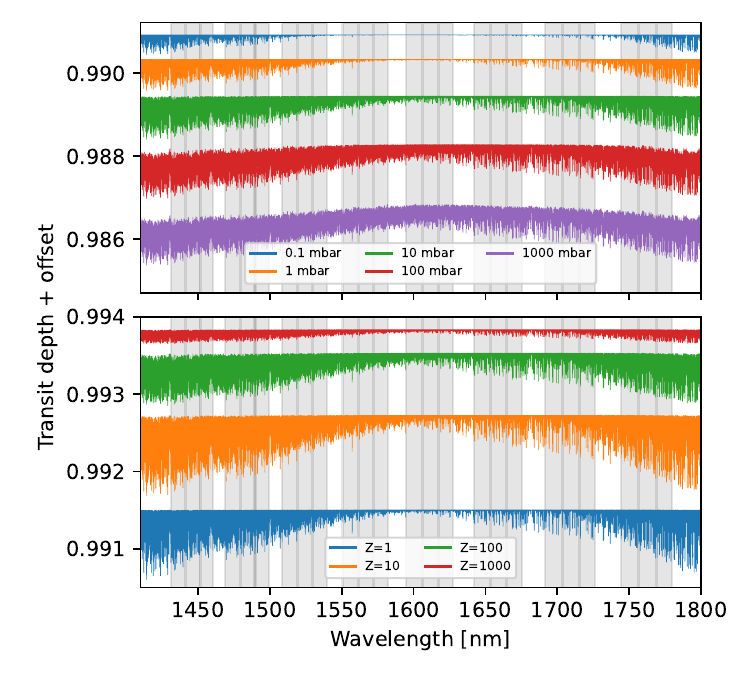}
    \caption{Examples of H$_2$O model atmospheres created with petitRADTRANS \citep{Molliere2019}. For clarity, an offset is added to the transit depth. The shaded regions illustrate the spectral coverage of CRIRES+. Top panel: 10$\times$ solar metallicity with varying cloud deck pressures. Bottom panel: Cloud deck at 10~mbar with varying metallicities, with $Z=1$ indicating a solar metallicity.}
    \label{fig:templates}
\end{figure}

Figure~\ref{fig:templates} shows some examples of our H$_2$O templates generated with pRT to illustrate the effects of varying cloud deck pressures and metallicities, convolved to the spectral resolution of CRIRES+. As shown in the top panel, spectra with clouds at lower pressures (higher altitudes) have more muted spectral features, which decreases their detectability. The spectral features also become weaker at very low and very high metallicities, as shown in the bottom panel. This is because at higher metallicites, the scale height of the atmosphere is reduced resulting in effectively smaller signals, while at lower metallicities, the spectral features get covered by the continuum produced by clouds and/or collision induced absorption.
%Very low metallicity atmospheres simply contain less of the targeted molecules resulting in weaker absorption. 
Nevertheless, even in these cases, absorption lines are potentially detectable at the high spectral resolution with CRIRES+. The strongest H$_2$O features can be seen in atmospheres with low-altitude or absent cloud decks with low to intermediate metallicites.

\subsection{Detection pipeline}\label{subsec:pipeline}

In our analysis, we make use of HRCCS for the detection of atmospheric species. The orbital parameters of GJ~436\,b found in the literature are relatively consistent. For the most accurate ephemeris, we assumed the period $P$ and transit midpoint $T_0$ as determined by the Transiting Exoplanet Survey Satellite (TESS, \citealt{TESS2015}). All of the parameters used in our analysis are listed in Table~\ref{tab:params}.

\begin{figure}
    \centering
    \includegraphics[width=\columnwidth]{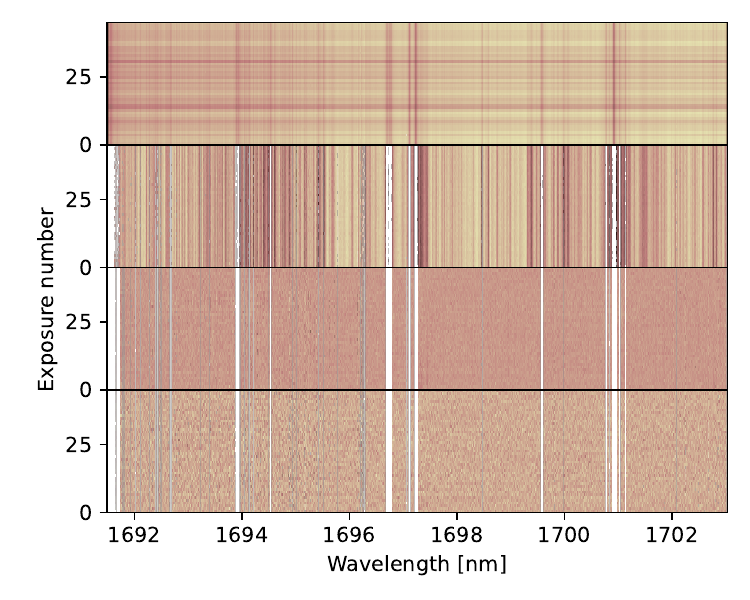}
    \caption{Stages of data processing shown for the second spectral order. Top panel: Observed spectra extracted from the CR2RES pipeline. Second panel: After masking of emission and absorption lines, continuum removal and outlier masking. Third panel: After removing the average off-transit spectrum. Bottom panel: After applying SYSREM.}
    \label{fig:process}
\end{figure}

Our pipeline consists of several steps, conducted for each nodding position separately,
%Firstly, the contaminating tellurics and stellar lines are removed. Subsequently, the residuals are cross-correlated with a template spectrum, namely the model atmospheres from Section~\ref{subsec:temp}. 
and is based on that of \cite{Landman2021}. Firstly, for each exposure and each spectral order, regions of the spectrum were masked where the flux is lower than 85\% and higher than 120\% of the normalized spectrum. This already removes a large fraction of emission and absorption lines caused by Earth's atmosphere, as well as some deep stellar lines. The continuum was normalized for each order separately by fitting a third degree polynomial and dividing by it. Outliers at three times the standard deviation were masked. The result of these steps can be seen in the second panel of Figure~\ref{fig:process}.

To correct for the still dominant telluric and stellar lines, each individual spectrum was divided by the average off-transit spectrum, with the result shown in the third panel of Figure~\ref{fig:process}. While this removed most of the remaining contaminating features, some telluric and stellar residuals remained. Therefore, the de-trending algorithm SYSREM, designed to remove systematic effects in a large set of lightcurves \citep{Tamuz2005,Mazeh2007}, was subsequently applied to each order separately. It is based on principle component analysis (PCA), but unlike PCA, SYSREM also considers the individual errors of each data point. Dividing by the average off-transit spectrum beforehand decreased the runtime of SYSREM considerably. After four SYSREM iterations, as determined to maximize the S/N of injected model atmospheres, a final removal of outliers, at more than three times the standard deviation, was applied. The resulting spectra visually resemble pure noise, as they should at this stage of processing, and are shown in the bottom panel of Figure~\ref{fig:process}.

Any potential planetary signal should be buried in the noise left over after removing the stellar and telluric features. Specific molecular species can now be searched for using cross-correlation with model spectra. Prior to cross-correlation, we Doppler-shifted the wavelengths of the data within a radial velocity range from -200 to +200 km/s with a step size of 1 km/s, and linearly interpolated the template spectrum onto the shifted wavelength grid. The cross-correlation was computed by taking the inner product of the error-weighted observed spectra and the shifted template spectra. This results in a 2D map, showing the cross-correlation strength as function of radial velocity and planet orbital phase. A signal in the cross-correlation map at the expected radial velocity of the planet would indicate a detection.

The exoplanet's radial velocity $v_\text{p}$ depending on its phase $\phi$, as seen from Earth's reference frame, can be calculated given its radial velocity semi-amplitude $K_\text{p}$, the barycentric velocity $v_{\text{bary}}$ caused by the Earth orbiting the barycenter of the Solar system, and the systemic velocity of the of the star–planet system $v_{\text{sys}}$:

\begin{equation}\label{equ:rv_pl}
    v_\text{p}(\phi)=K_\text{p} \cdot \text{sin}(2\pi \phi) - v_{\text{bary}} + v_{\text{sys}}
\end{equation}

The barycentric velocity $v_{\text{bary}}$ is obtained from the time information of our data. The radial velocity semi-amplitude $K_\text{p} $ is calculated using the semi-major axis $a$ (derived from the planet mass $M_\text{p}$ and stellar mass $M_\text{s}$), period $P$, eccentricity $e$ and inclination $i$: 

\begin{equation} \label{equ:kp}
    K_\text{p} =\frac{2\pi a}{P\sqrt{1-e^2}} \text{sin}(i)  \quad \text{, with} \,\,a =\left( \frac{G\cdot (M_\text{s}- M_\text{p})\cdot P^2}{4\pi^2}\right)^{1/3}
\end{equation}

The semi-amplitude of a circular orbit is $K_\text{p}(e=0)=K_\text{p,c}=117.03$~km/s, while including the eccentricity results in a marginally larger value of $K_\text{p}(e=0.145)=K_\text{p,e}=118.28$~km/s. However, accounting for the eccentricity only in $K_\text{p}$ assumes the standard convention that the planetary argument of periastron is $\omega_\text{p}=270^{\circ}$ at the middle of the transit (i.e., at an orbital phase of zero). Since GJ~436\,b has an argument of periastron of $\omega_\text{p}=145.8^{\circ}$, the above equations are insufficient for describing the planet's orbit accurately. For eccentric orbits, the value of $\omega_\text{p}$ can have a profound impact on the orbital configuration and hence the radial velocity of the planet. Therefore, the eccentricity $e$, the planetary argument of periastron $\omega_\text{p}$ and the true anomaly $\nu(\phi)$ must be taken into account (e.g., \citealt{Wright2009}, \citealt{Basilicata2024}) when calculating the planet's radial velocity $v_\text{p}$:

\begin{equation}\label{equ:rv_pl_e}
    v_\text{p}(\phi)=K_\text{p} \cdot [\text{cos}(\nu(\phi)+\omega_\text{p})+e\cdot\text{cos}(\omega_\text{p})]- v_{\text{bary}} + v_{\text{sys}}
\end{equation}

We used the python package ``kepler.py"\footnote{\url{https://pypi.org/project/kepler.py/}} to derive the true anomaly $\nu(\phi)$ as a function of the phase, based on the planetary argument of periastron. To highlight the impact of the eccentricity on $v_\text{p}$, we calculated the planet's radial velocity for a circular (Equ.~\ref{equ:rv_pl} and Equ.~\ref{equ:kp} with $e=0$) and eccentric (Equ.~\ref{equ:rv_pl_e} and Equ.~\ref{equ:kp} with $e=0.145$) orbit for a direct comparison, as shown in Figure~\ref{fig:CCF_Kpv}.

\begin{table}[t!]
    \caption{Parameters of the GJ 436 system used in our analysis.}
    \centering
    \begin{tabular}{lll}
    \hline
    \hline
    Parameter & Value & Ref \\
    \hline
    %\textbf{Planetary} & & \\
    $P$ [days] & $2.64388561055203\pm0.00003158837$ & 1 \\
    $T_0$ [BJD] & $2459639.96333\pm0.00018952927$ & 1 \\
    $T_d$ [hours] & $1.009\pm0.034$ & 2\\
    %$a$ [AU] & $0.02849\pm0.00020$ & 3\\
    $e$ & $0.145\pm0.027$ & 3\\
    $i$ [deg]& $86.858\substack{+0.049\\-0.052}$ &4\\
    $R_{\text{p}}$ [R$_{\text{Jup}}$]  & $0.3739\pm0.0097$ &5\\
    $M_{\text{p}}$ [M$_{\text{Jup}}$]  & $0.0669\pm0.0022 $ & 3\\
    $R_{\text{s}}$ [R$_{\odot}$] & $0.416843\pm0.0075623$ &3\\
    $M_{\text{s}}$ [M$_{\odot}$] & $0.4411671425335\pm0.00938690255671 $ &3\\
    $v_{\text{sys}}$ [km/s] & $9.59\pm0.8$ &6\\
    $T_\text{eq}$ [K]& $686\pm10$&5\\
    $\omega_\text{p}$ [deg] & $145.8\substack{+5.4\\-5.7}$& 7 \\    
         \hline
    \end{tabular}
    \begin{tablenotes}
      \small
      \item \textbf{Notes.} We use the period $P$, transit midpoint $T_0$, transit duration $T_d$, eccentricity $e$, inclination $i$, planetary radius $R_{\text{p}}$, planetary mass $M_{\text{p}}$, stellar radius $R_{\text{s}}$, stellar mass $M_{\text{s}}$, system radial velocity $v_{\text{sys}}$, planetary equilibrium temperature $T_\text{eq}$, and planetary argument of periastron $\omega_\text{p}$ from the following references, numbered in the third column: 1) \cite{TESS2015}, 2) \cite{Baluev2015}, 3) \cite{Rosenthal2021}, 4) \cite{Bourrier2018}, 5) \cite{Turner2016},
      6) \cite{Fouqu2018}, 7) \cite{Trifonov2018}. 
    \end{tablenotes}
    \label{tab:params}
\end{table}

To combine the signals from all in-transit observations, the cross-correlation functions (CCFs) were shifted to the rest frame of the GJ~436 system for each in-transit exposure and were summed. The CCFs were shifted according to the radial velocity from Equation~\ref{equ:rv_pl} using $K_\text{p}$ values ranging from -250 to +350 km/s, to check if the strongest signal appears at the expected $K_\text{p}$. The value of $K_\text{p}$ reflects the radial velocity change during the transit, but due to the distortion of the orbit through the eccentricity, the effective $K_\text{p,eff}$ for the eccentric orbit is not identical to $K_\text{p}(e=0.145)=K_\text{p,e}$ as calculated through Equ.~\ref{equ:kp}. Instead, it is determined through the circular $K_\text{p,c}$ (from Equ.~\ref{equ:kp} with $e=0$) and the ratio of radial velocity change for the eccentric ($\Delta v_\text{p,e}$) and circular ($\Delta v_\text{p,c}$) orbit during the transit:

\begin{equation}
\frac{K_\text{p,eff}}{K_\text{p,c}} = \frac{\Delta v_\text{p,e}}{\Delta v_\text{p,c}}
\end{equation}

The equation above results in an effective semi-amplitude of $K_\text{p,eff}=102.25$~km/s, which is where one would expect to find a planetary signal in the case of a detection. All steps until this point were run for both nodding positions, but now combined  to increase any potential signal. The S/N of the cross-correlation peak was estimated by dividing the signal by the standard deviation, which was calculated excluding the velocity region of $\pm40$~km/s around the a system rest frame velocity of zero, where one would expect the signal to be in the case of a circular orbit, and slightly shifted in the case of an eccentric orbit.

Our detection pipeline was calibrated by injecting a model spectrum at the planet's radial velocity into the in-transit exposures. For this, we used a H$_2$O template with a solar metallicity and cloud deck at 10~mbar. The S/N of the injected model was determined by running the pipeline with and without the injection and subsequently taking the difference of the CCFs, which we then divided by the standard deviation of the non-injected run. As such, the model spectrum's retrieved S/N is the resulting value at the planet's rest frame velocity and $K_\text{p}$. Through this process, we obtained the above mentioned optimum of four SYSREM iterations and removal of outliers at a standard deviation larger than three. In this way, we also found that employing a weighting scheme for the different orders based on the recovered S/N of the injected model does not lead to a higher S/N when removing the telluric-dominated orders from the analysis. In fact, only discarding the three most telluric-dominated orders without weighting produced the highest S/N of the recovered injected signal. Therefore, we constructed our pipeline accordingly. Both nodding positions yielded similar results. 

\section{Results}\label{sec:results}

\begin{figure}
    \centering
    \includegraphics[width=\linewidth]{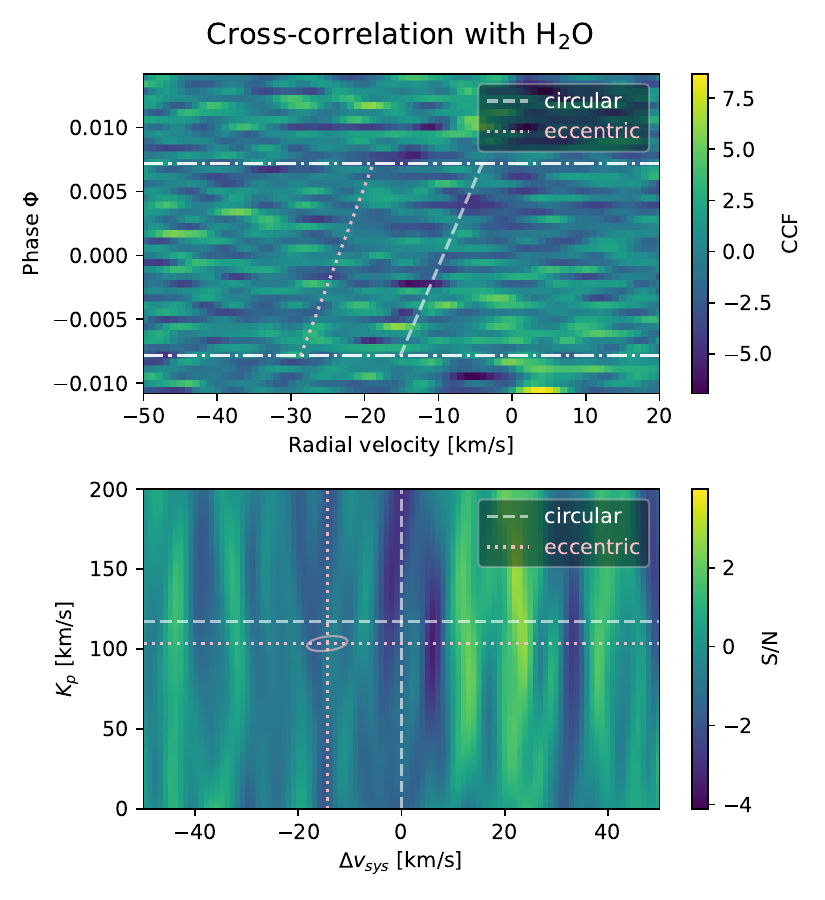}
    \caption{Cross-correlation strength from the H$_2$O template with a cloud deck at 10~mbar and 10$\times$ solar metallicity, summed over all spectral orders and both nodding positions. Top panel: CCFs for each spectrum, with the radial velocity in Earth's rest frame on the x-axis and the transit phase and corresponding exposure number on the y-axes. The horizontal dashed lines indicate the beginning and end of the transit, while the slanted lines (circular orbit in white, true eccentric orbit in pink) shows the radial velocity of the planet during its transit and hence the expected location of the signal. Bottom panel: Estimated S/N as a function of the planetary rest-frame velocity $v_\text{sys}$ and radial velocity semi-amplitude $K_\text{p}$. The intersection of the vertical and horizontal dashed lines (circular orbit in white, true eccentric orbit in pink) indicate the location of the expected signal in the absence of atmospheric winds. The pink ellipse encompasses the $K_\text{p}$ and radial velocity values when considering the 1-$\sigma$ errors of  $\omega_\text{p}, e,$ and $M_\text{s}$.}
    \label{fig:CCF_Kpv}
\end{figure}

The detection pipeline was run using the generated H$_2$O, CH$_4$, and CO model atmospheres as templates for the cross-correlation. No signal was detected from any of these molecules in the data. Cross-correlating with OH and NH$_3$ model atmospheres also resulted in non-detections.

As an example, we show the cross-correlation results from the H$_2$O template with a cloud deck at 10~mbar and 10$\times$ solar metallicity. The top panel of Figure~\ref{fig:CCF_Kpv} shows the cross-correlation strength for each spectrum (both nodding positions combined), summed over all spectral orders, as a function of the radial velocity and transit phase. The transit occurs between the horizontal dashed lines. The expected location of the planetary signal, assuming the true eccentric orbit or a circular orbit, are shown by slanted lines in pink and white, respectively. Accounting for the altered orbital configuration due to the eccentricity results in a significantly different radial velocity of the planet during its transit, shifted by an average of -14.3 km/s with respect to the circular orbit. The slope of the lines also differ slightly, reflecting a change in $K_\text{p}$ from $K_\text{p,c}=117.03$~km/s (circular) to $K_\text{p,eff}=102.25$~km/s (effective eccentric). This highlights the need for accurate orbital calculations, which incorporate the eccentricity and argument of periastron, in studies that depend on the planet's radial velocity. Precise values of these parameters are crucial for a correct orbital solution, as explored for instance by \cite{Basilicata2024}. Failing to account for the effect of the eccentricity can, at best, partially smear the planetary signal, and at worst, shift it to completely different values and hinder a potential detection. 

The bottom panel of Figure~\ref{fig:CCF_Kpv} shows the estimated S/N derived from the CCFs, representing the detection significance, as a function of the system rest-frame velocity $v_\text{sys}$ and radial velocity semi-amplitude $K_\text{p}$. The intersection of the dashed lines show the expected location for the true eccentric orbit (pink) and circular orbit (white). In the rest-frame of the star, a planet on a circular orbit would have a radial velocity of zero during the transit, hence the position of the circular orbit at $v_\text{sys}=0$~km/s and at $K_\text{p,c}=117.03$~km/s. In contrast, the eccentric orbit of the planet leads to a shifted position in the $K_\text{p}-v_\text{sys}$ diagram at $v_\text{sys}=-14.3$~km/s and at $K_\text{p,eff}=102.25$~km/s. In the case of a detection, a signal would appear where the pink lines intersect. No signal is detected at the expected location or anywhere else in the examined velocity range. In order to assess the impact of parameter uncertainties on $K_\text{p,eff}$ and $v_\text{p}$, we drew 1000 random values for $e$, $\omega_\text{p}$, and $M_{\text{s}}$ from a uniform distribution of spanning $\pm 1\sigma$, from which we calculated $K_\text{p,eff}$ and $v_\text{p}$. The resulting range is highlighted by the pink ellipse. We conclude that the effect of the uncertainties is not too large, affecting $K_\text{p,eff}$ by roughly $\pm4$~km/s and $v_\text{p}$ by $\pm5$~km/s.

%In the case of a detection, a signal would appear at $v_\text{sys}=0$km/s and $K_\text{p}=118.28$km/s, where the dashed lines intersect. 

\begin{figure*}
    \centering
    \includegraphics[width=\textwidth]{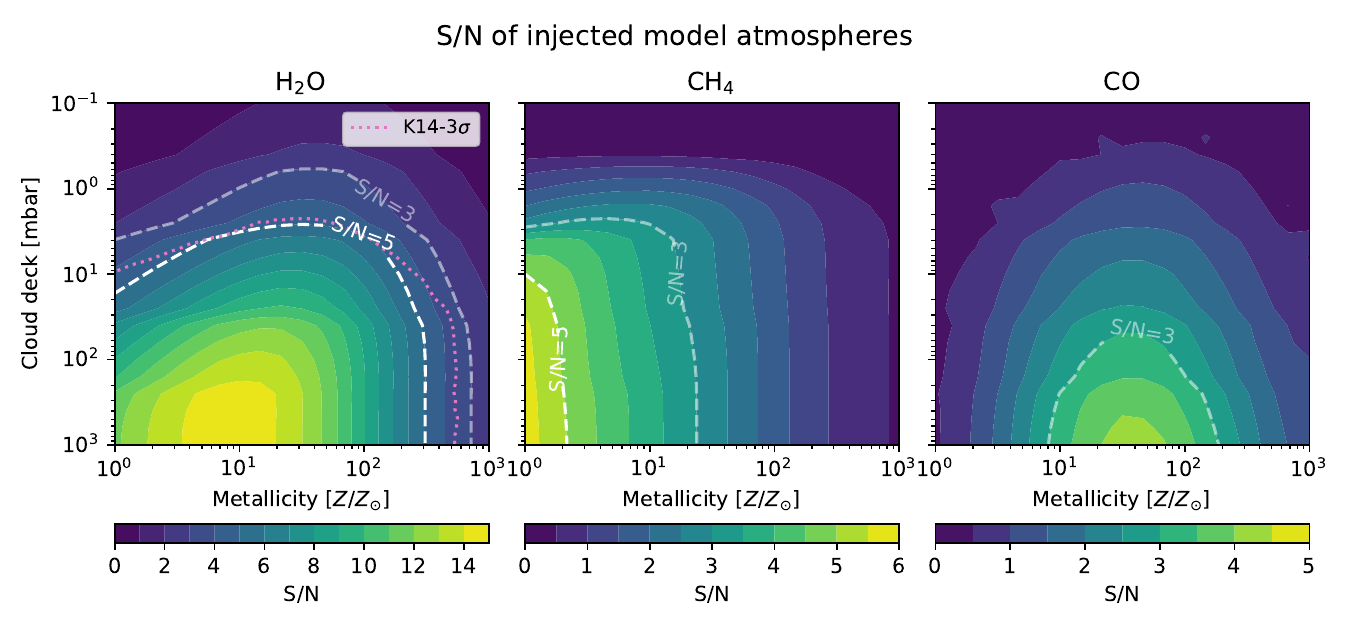}
    \caption{Retrieved S/N of the injected model atmospheres for H$_2$O (left), CH$_4$ (middle), and CO (right) with varying metallicites and cloud deck altitudes. The white dashed lines indicate the contours for S/N=3 and S/N=5, with the latter taken as a detection threshold. The pink dotted curve (labeled as K14-$3\sigma$) is the 99.7\% constraint from Fig.~3 in \cite{Knutson2014}.}
    \label{fig:cloud_metall_2D}
\end{figure*}

\begin{figure*}
    \centering
    \includegraphics[width=\textwidth]{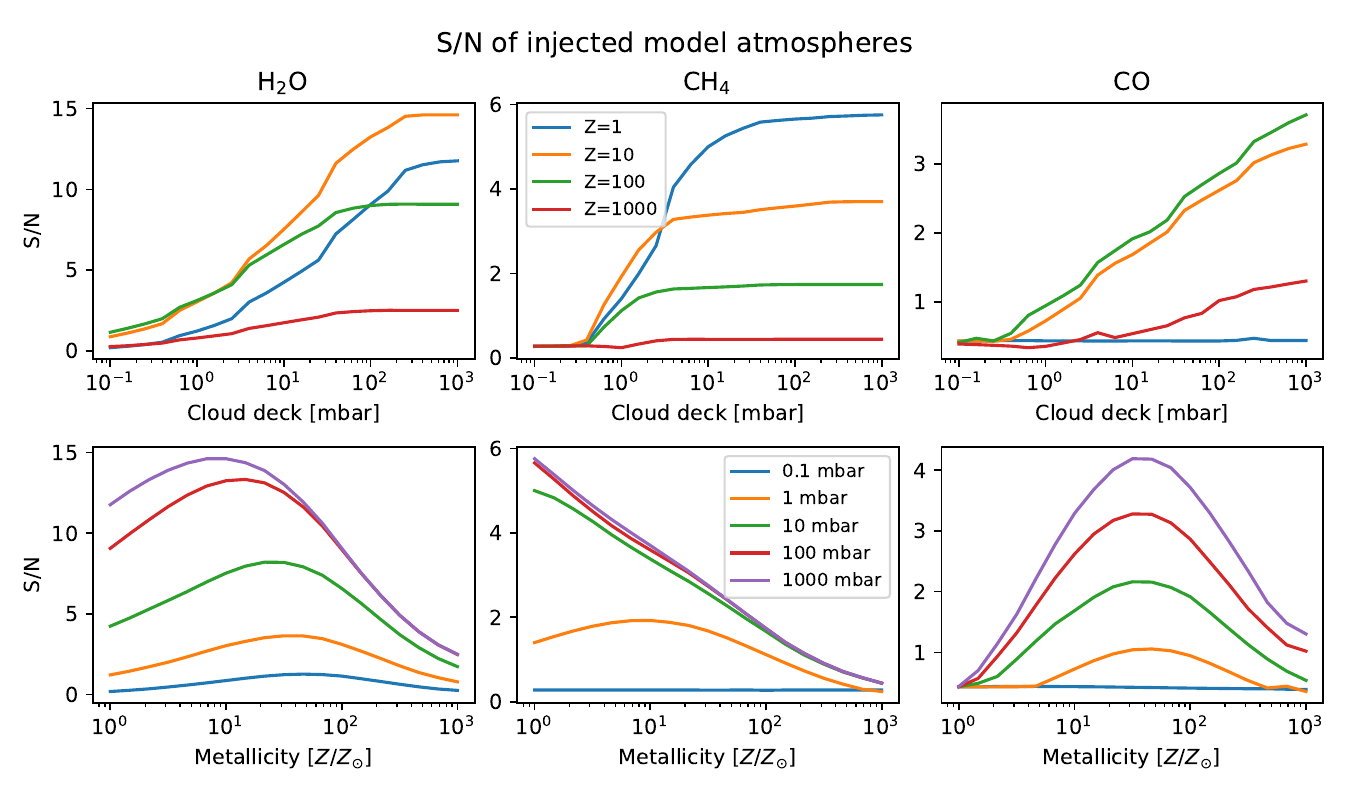}
    \caption{Retrieved S/N of injected model atmospheres for H$_2$O (left column), CH$_4$ (middle column), and CO (right column), showing slices of selected metallicities $Z$ (top row) and cloud deck altitudes (bottom row) from Figure~\ref{fig:cloud_metall_2D}.}
    \label{fig:cloud_metall_slices}
\end{figure*}

We use the data to constrain the atmospheric composition of GJ~436\,b by determining the retrievability of various injected model atmospheres. For this, we use our previously generated grid of H$_2$O, CH$_4$, and CO templates with varying metallicities, ranging from solar to 1000 times solar, and cloud deck altitudes, ranging from 0.1~mbar to 1 bar. We ran our pipeline on each injected model atmosphere and determined the resulting S/N.

Figure~\ref{fig:cloud_metall_2D} shows a 2D map of the retrieved S/N of the injected models for H$_2$O (left), CH$_4$ (middle), and CO (right) depending on metallicity and pressure. We highlight the contours of S/N=3 and S/N=5, and consider the latter as a detection threshold. For a comparison with the constraints from the HST data, the 99.7\% posterior probability density from Fig.~3 in \cite{Knutson2014} is included. Due to the unavailability of the contour source data, these values were extracted from their figure using WebPlotDigitizer (Version 4.6, \citealt{Rohatgi2022}).

Any injected model atmosphere with a recovered S/N>5, as shown in Figure~\ref{fig:cloud_metall_2D}, can be considered a detection, and can be ruled out as a possible atmosphere of GJ~436\,b. Due to the highest achievable S/N, if present, H$_2$O should be the most readily detectable molecule in this data. This is unsurprising, given the considerably larger transmission signal produced by H$_2$O compared to CH$_4$ and CO. Furthermore, in the wavelength range covered by CRIRES+, CO spectral features are only present in a very narrow range. Similarly, OH only has a few strong spectral lines in our considered wavelength range, while NH$_3$ produces only a relatively small transmission signal.

Given our non-detection, we can rule out atmospheres in the bottom left corner of the H$_2$O plot in Figure~\ref{fig:cloud_metall_2D}, that is, atmospheres with low-altitude cloud decks ($P$>10~mbar) and simultaneous solar to 300$\times$ solar metallicities. Our analysis indicates that the most likely scenario for GJ~436\,b is a high-altitude cloud deck ($P$<10~mbar) and/or a high metallicity (>300$\times$ solar), since it falls beyond the scope of what is detectable with the data. Given the planet radius and mass, a solar metallicity is less likely, but intermediate metallicities are expected \citep{Figueira2009}.

Figure~\ref{fig:cloud_metall_slices} shows slices from Figure~\ref{fig:cloud_metall_2D} for selected metallicities (top row) and cloud decks (bottom row). In the case of all three molecules, a cloud deck at high pressures, and hence low altitudes, enhances the retrieved S/N, as seen in the top row of Figure~\ref{fig:cloud_metall_slices}. Atmospheres with cloud decks at very low altitudes and hence high pressures are equivalent to cloud-free atmospheres. However, the effect of metallicity varies depending on the examined molecule, as seen in the bottom row of Figure~\ref{fig:cloud_metall_slices}. The metallicity of an atmosphere can impact the retrieved S/N in two ways. For one, an atmosphere with a higher metallicity can increase the S/N due to containing more of the targeted molecules, which together produce stronger spectral features that are more prominent with respect to the continuum. On the other hand, a larger fraction of heavy elements also decreases the scale height of the atmosphere, which has the opposite effect.

Atmospheres in chemical equilibrium with temperatures like that of GJ~436\,b are expected to contain H$_2$O and CH$_4$ due to their favored formation pathways, governed mainly by the reaction CO+3H$_2$ $\rightleftharpoons$ CH$_4$+H$_2$O. At lower temperatures and higher pressures, the equilibrium shifts towards the formation of CH$_4$ \citep{Seager2010}. However, as cross-correlation is considerably more sensitive to line positions than their depths and hence absolute absolute abundances, a higher S/N does not necessarily reflect a higher abundance and may instead be due to the higher number and density of lines in the probed wavelength region, resulting in a higher correlation peak. This can be seen in Figure~\ref{fig:cloud_metall_2D}, where the higher obtainable S/N of the H$_2$O signal arises due to its denser forest of lines in the H-band, while the relatively few spectral features of CO account for the lower S/N.

For H$_2$O, a 10$\times$ solar metallicity produces the highest S/N. In Figure~\ref{fig:cloud_metall_slices}, one can also see that for higher metallicities the S/N remains flat over a larger range of cloud deck altitudes. This is because the signal is mainly coming from higher in the atmosphere, so the cloud deck only starts to dampen the signal once it is at high altitudes. For CH$_4$, a solar metallicity produces the highest S/N, while for CO, the highest S/N is achieved at about 50$\times$ solar metallicity. In our model atmospheres, assuming chemical equilibrium, the mass fractions of CH$_4$ and CO both peak for higher metallicities (100--1000$\times$ solar), but this does not reflect in the retrieved S/N of the injected signals, as cross-correlation is not very sensitive to absolute abundances. For CH$_4$, with prominent spectral features throughout our probed wavelength range, the decreasing scale height with higher metallicities appears to be producing the largest effect on its detectability. In contrast, the spectral features of CO are only present in a very narrow wavelength range and much less prominent with respect to the continuum at very high or low metallicities. Despite our non-detections, the presence of these molecules in GJ~436\,b's atmosphere cannot be ruled out and requires further investigations.

%For CH$_4$, a solar metallicity is more favorable, due to the molecule having four hydrogen atoms and thus requiring a higher abundance of light elements. As CO only consists of heavier elements, its highest S/N is achieved at about 50$\times$ solar metallicity, consequently higher than that needed for H$_2$O.

\section{Discussion}\label{sec:discuss}

Our results are consistent with the flat transmission spectrum found by \cite{Knutson2014}, produced either by high-altitude clouds and/or a high atmospheric metallicity. Assuming Gaussian noise properties (e.g., S/N=3 is comparable to a probability of 99.7\%), we compare the upper limits from Figure~\ref{fig:cloud_metall_2D} to the constraints shown in their Figure~3, which are derived from four transits of HST data. Comparing our constraints from the S/N to the posterior probability density of \cite{Knutson2014} is of course a simplified approach and only serves as an approximation. However, since we only deal with injected signals, for which we know the position and line-shape exactly, this significantly reduces any biases. %The advantage of using the S/N of the injected models after cross-correlation as a comparison is that it accounts for systematic effects and correlated noise (such as residuals from telluric water lines), while likelihood-based approaches assume uncorrelated data points.

The highlighted contours in Figure~\ref{fig:cloud_metall_2D} show that the $3\sigma$ constraints derived by \cite{Knutson2014} fall between our S/N=3 and S/N=5 limits. Consequently, for a given metallicity and cloud deck pressure, we estimate that CRIRES+ data would provide a somewhat higher S/N. From this, we postulate that one transit with CRIRES+ can offer constraints of slightly better or similar quality compared to four transits with HST. Furthermore, our findings also agree with other modeling and observational studies \citep{Lanotte2014,Agundez2014,Morley2017,Lothringer2018} that are compatible with a high-metallicity atmosphere. 

The detection of H$_2$O may be impeded by Earth's telluric absorption, which overlaps with some strong H$_2$O spectral features, in particular around $1400$~nm. In the case of cloudy atmospheres, the signal from above the clouds comes mainly from the wavelength regions that also harbor strong telluric features. The fact that GJ~436 is an M-dwarf which also has H$_2$O features further complicates the detection. The planet's small radial velocity change during the transit makes it more difficult to disentangle its H$_2$O features from those of the star.

%Moreover, GJ~436\,b is a cooler planet compared to the many hot Jupiters targeted so far with HRCCS, further making its signal weaker and harder to detect. Since GJ~436\,b most likely has either a high-altitude cloud deck or a high-metallicity atmosphere, it is a particularly challenging target for observations. 
 
Our non-detection is probably mainly due to the fact that we only obtained good quality data from one observing night. An additional two to three transits of high S/N might be sufficient for a detection, assuming that the S/N increases with the square root of the number of observations. The required number of transits for detection of a certain atmosphere can be estimated based on the S/N of the retrieved injected signals in Figure~\ref{fig:cloud_metall_2D}. Given that we would obtain a S/N$\sim$3 for a 10--100$\times$ solar metallicity atmosphere with a cloud deck at 1~mbar, only three observations of similar high quality would be required to reach a detection threshold of $\text{S/N}\sim5$. Similarly, we would detect a high-metallicity atmosphere (1000$\times$ solar) with a low-altitude cloud deck ($P$>100~mbar) at a S/N$\sim$2.5, so only four high quality transits would be required to reach S/N$\sim$5. Only the detection of high-metallicity atmospheres together with a high-altitude cloud deck would require an unfeasible amount of transits with CRIRES+, such as 40 transits for a 1000$\times$ solar metallicity and cloud deck at 1~mbar. In this regard, despite the promising potential of HRCCS for characterizing exoplanet atmospheres and higher sensitivity compared to HST, the combination of several high-quality observing nights may be required in the case of cloudy and/or high metallicity atmospheres. 

\cite{Gandhi2020} predicted that HRS on 4~m class telescopes, such as SPIRou, CARMENES or GIANO, could partially break the degeneracy between high metallicities and high-altitude clouds using 10~hours of transit observations, which amounts to 10 transits in total given the planet's 1~hour transit duration. A direct comparison with their study is difficult due to significant methodological differences. For one, they deal with other instruments, which have a larger and more continuous wavelength coverage than CRIRES+, but are on smaller telescopes. Considering our above extrapolation from the S/N of injected model atmospheres, which shows that only four high quality transits would be required for a S/N$\sim$5, combined with a twice as large telescope diameter, our findings appear to be more or less consistent with the predictions by \cite{Gandhi2020}.

Alternatively, observations with other facilities also hold potential, such as the near-infrared spectrometer (NIRSpec) on the James Webb Space Telescope (JWST). \cite{Constantinou2022} find that single-transit transmission spectra combining all three NIRSpec gratings can constrain H$_2$O, CH$_4$, and NH$_3$ of the two temperate mini-Neptunes K2-18\,b and TOI-732\,c, for cloud deck pressures as low as 3 and 0.1~mbar, respectively, in the case of a 10$\times$ solar metallicity. In fact, the potential of JWST for characterizing sub-Neptunes is already being demonstrated by unprecedented findings, such as the detections of CH$_4$ and CO$_2$ in the mini-Neptune K2-18\,b \citep{Madhusudhan2023} and in the sub-Neptune TOI-270\,d \citep{Holmberg2024}. However, the transmission spectrum of the sub-Neptune TOI-836\,c also appears to be featureless with JWST data \citep{Wallack2024}.

\section{Conclusion} \label{sec:concl}

We have obtained new high-resolution CRIRES+ observations of the warm Neptune GJ~436\,b. Cross-correlating the data with H$_2$O, CH$_4$, and CO model atmospheres of various metallicities and cloud deck altitudes yields a non-detection. Through injecting model atmospheres into the data and determining their retrievability, a range of atmospheric compositions can be ruled out, such as low to intermediate metallicities combined with a low-altitude cloud deck or a cloud-free atmosphere. Our analysis agrees with previous findings that the atmosphere of GJ~436\,b most likely has a high-altitude cloud deck ($P$<10~mbar) and/or a high-metallicity (>300$\times$ solar). Finally, a comparison of our detection thresholds with those derived from HST data \citep{Knutson2014} appear to yield slightly better constraints. Therefore, we estimate that one transit with CRIRES+ may be somewhat higher in quality compared to four transits with HST. While this illustrates the promising potential of CRIRES+ spectroscopy for characterizing exoplanet atmospheres, the combination of at least several high S/N transits will be necessary for the detection of chemical species in cloudy and/or high metallicity atmospheres such as that of GJ~436\,b.

\begin{acknowledgements}
We extend our gratitude to the anonymous referee, whose constructive feedback has significantly helped improve our paper. N.G. also wishes to thank M. Basilicata for the very helpful explanations about calculating eccentric orbits. Support for this work was provided by NL-NWO Spinoza SPI.2022.004. Our work is based on observations collected at the European Organisation for Astronomical Research in the Southern Hemisphere under ESO programme 110.2492.004. D.G.P acknowledges support from NWO grant OCENW.M.21.010.
\end{acknowledgements}

\bibliographystyle{aa}
\bibliography{GJ436b.bib}

\end{document}